\documentclass[twocolumn]{aastex631} 

\usepackage{graphicx} 
\usepackage{amsmath}
\usepackage{amssymb}
\usepackage{bm}
\usepackage{natbib}
\usepackage{hyperref}
\hypersetup{
    colorlinks = true,
    citecolor  = blue,
    linkcolor  = blue,
    urlcolor   = blue
}
\usepackage{csquotes} 

\newcommand{\params}{\boldsymbol{\theta}}
\newcommand\prob[1]{p(#1)}

\shorttitle{Bayesian Abundance Inference}
\shortauthors{Kuske, Arcones, Svensson}

\begin{document}

\title{Bayesian Inference of Stellar r-Process Abundances}

\author[0009-0005-5121-7343]{Jan Kuske}
\affiliation{Institut für Kernphysik, Technische Universität Darmstadt, Schlossgartenstr. 2, Darmstadt 64289, Germany}

\author[0000-0002-6995-3032]{Almudena Arcones}
\affiliation{Institut für Kernphysik, Technische Universität Darmstadt, Schlossgartenstr. 2, Darmstadt 64289, Germany}
\affiliation{GSI Helmholtzzentrum für Schwerionenforschung GmbH, Planckstr. 1, Darmstadt 64291, Germany}
\affiliation{Max-Planck-Institut für Kernphysik, Saupfercheckweg 1, 69117 Heidelberg, Germany}

\author[0000-0002-9211-5555]{Isak Svensson}
\affiliation{Institut für Kernphysik, Technische Universität Darmstadt, Schlossgartenstr. 2, Darmstadt 64289, Germany}
\affiliation{GSI Helmholtzzentrum für Schwerionenforschung GmbH, Planckstr. 1, Darmstadt 64291, Germany}
\affiliation{Max-Planck-Institut für Kernphysik, Saupfercheckweg 1, 69117 Heidelberg, Germany}

\begin{abstract}
The abundances of heavy elements observed in metal-poor, r-process-enhanced stars provide unique information about the astrophysical conditions in which the rapid neutron-capture process (r-process) occurs. We present a Bayesian framework to infer these conditions from stellar abundance patterns. Building on a large site-independent r-process survey, we use a Markov chain Monte Carlo (MCMC) sampler to fit observed abundances with weighted superpositions of abundances from nucleosynthesis calculations, each parameterized by an initial electron fraction, entropy, and expansion timescale. Applying this framework to the nearly complete r-process template star HD222925, we find that two components are required: A heavier H-component producing elements from the second to the third peak and a lighter L-component producing elements from the first to the second peak. Increasing the number of components does not significantly improve the agreement with observations. Extending the analysis to the limited-r stars HD128279 and HD122563, and to the solar r-process residuals, we find that their abundances can be reproduced by almost the same two components, albeit with different relative weights. Moreover, the lightest neutron-capture elements ($Z \lesssim 35$) in the Sun require additional contributions absent in HD222925. Residual discrepancies are concentrated in elements sensitive to observational systematics (e.g., Ag and Cd) or nuclear-physics inputs (around the third peak). Our study highlights the power of combining stellar abundances with nucleosynthesis calculations to constrain the astrophysical conditions of the r-process, and motivates further improvements in both observational data and nuclear-physics inputs.
\end{abstract}


\section{Introduction}
\label{sec:intro}

Abundances derived from the spectra of metal-poor, r-process-enhanced stars are a key observable for understanding rapid neutron-capture (r-process) nucleosynthesis. Because these old stars formed from gas polluted by only a few nucleosynthetic events, their abundance patterns preserve a relatively clean imprint of individual r-process events. Complementary, the solar r-process residuals can be obtained by subtracting the calculated s-process contribution from the total solar abundances \citep[e.g.,][]{Goriely1999_UncertaintiesSolarSystem, Sneden2008_NeutronCaptureElementsEarly, Prantzos2020_ChemicalEvolutionRotating}.

The pattern (relative abundance) of elements between the second ($A \approx 130$) and third ($A \approx 195$) r-process peaks is remarkably robust across stars and in the solar r-process residuals \citep[e.g.,][]{Sneden2008_NeutronCaptureElementsEarly, Racca2025_RProcessAllianceExploringa}. This robustness is  in contrast to the lighter heavy elements (Sr, Y, Zr up to the second peak) that show considerable star-to-star variability \citep[e.g.,][but also \citealp{Roederer2022_RProcessAllianceAbundance}]{Roederer2010_UbiquityRapidNeutroncapture, Hansen2014_HowManyNucleosynthesis, Spite2018_AbundancePatternsLight}. Suggestions to explain this variability include a lighter element primary process \citep[e.g.,][]{Travaglio2004_GalacticEvolutionSr, Francois2007_FirstStarsVIII, Montes2007_NucleosynthesisEarlyGalaxy}, fission \citep{Roederer2023_ElementAbundancePatterns}, weak r- or $\nu$p-processes \citep[e.g.,][]{Frohlich2006_NeutrinoinducedNucleosynthesis64, Arcones2011_ProductionLightelementPrimary, Bliss2018_SurveyAstrophysicalConditions, Psaltis2024_NeutrinodrivenOutflowsElemental}, an $\alpha$-process \citep{Woosley1992_AlphaProcessRProcess}, and contributions from spinstars \citep{Frischknecht2012_NonstandardSprocessLow}, among others. The first-peak region ($A \approx 80$) in particular reveals notable differences between solar r-residuals and metal-poor stars, pointing to additional production channels. Moreover, Th and U are enhanced in some so-called actinide-boost stars \citep[e.g.,][]{Schatz2002_ThoriumUraniumChronometers, Roederer2009_EndNucleosynthesisProduction, Holmbeck2018_RProcessAlliance2MASS}.

Previous works have fitted observed abundances with nucleosynthesis calculations; e.g., \citet{Kratz1993_IsotopicRProcessAbundances, Goriely1996_WaitingPointApproximation} and \citet{Freiburghaus1999_AstrophysicalRprocessComparison} concluded that a superposition of at least three different astrophysical conditions is required to reproduce the solar r-process abundances. Other works \citep[e.g.,][]{Qian2001_ModelAbundancesMetalpoor, Qian2007_WhereOhWhere, Hansen2014_HowManyNucleosynthesis} reversed the approach by starting from stellar abundance patterns and separating them into individual nucleosynthesis components. \citet{Holmbeck2019_ActiniderichActinidepoorRprocessenhanced, Holmbeck2023_HD222925New} used a Monte Carlo approach to fit the abundances of r-process-enhanced stars by varying the initial electron fraction in nucleosynthesis calculations of two representative trajectories. These studies highlight the importance of combining observational data with theoretical models to constrain the conditions under which the r-process operates.

In this work, we develop a Bayesian inference framework that uses a Markov chain Monte Carlo (MCMC) sampling to fit observed stellar abundance patterns with superpositions of nucleosynthesis calculations obtained from our large site-independent survey \citep[][hereafter KAR25]{Kuske2025_CompleteSurveyRprocess}. This approach simultaneously determines which combinations of astrophysical conditions reproduce a given star, the number of components required, and the associated uncertainties. We further introduce two complementary priors---an agnostic uniform prior and a simulation-informed prior built from a large compilation of hydrodynamical trajectories---allowing us to assess whether the conditions favored by  observations are also commonly found in current simulations. We apply the method to the nearly complete r-process template star HD222925, to a set of limited-r stars, and to the solar r-process residuals, and discuss the implications for the number of contributing components and for the origin of the remaining discrepancies.

The paper is organized as follows. In Section~\ref{sec:method} we describe the parametric nucleosynthesis calculations, the observational data, and the statistical inference setup, including the construction of the two priors. In Section~\ref{sec:results} we present the results for HD222925, study the dependence on the number of components, and compare different stars and the solar r-process. We summarize our findings and discuss open questions in Section~\ref{sec:conclusions}.

\section{Methods}
\label{sec:method}

\subsection{Parametric astrophysical conditions, nucleosynthesis and observations}
\label{sec:param}

This work is based on our large site-independent r-process survey published in KAR25. In that survey, state-of-the-art nucleosynthesis calculations were performed for a wide set of conditions, using the nuclear reaction network \textsc{WinNet} \citep{Reichert2023_NuclearReactionNetwork}. The 120\,000 different conditions cover those found in current hydrodynamical simulations and also go beyond these. The model uses the parametric density profile of \citet{Lippuner2015_RprocessLanthanideProduction} with three parameters: initial electron fraction $Y_\mathrm{e}$, initial entropy $s$, and expansion timescale $\tau$. Most nuclear physics inputs are based on the FRDM2012 mass model, see KAR25 for more details.

We compare calculated and observed abundances to constrain the number of contributing patterns (or components) to the r-process and to determine their conditions. We already showed in KAR25 that the abundances from single conditions (i.e., one combination of $Y_\mathrm{e}$, $s$, and $\tau$) cannot explain the full r-process pattern as observed in metal-poor stars and the solar r-process residual abundances. Instead, two or three different conditions have to be combined to reproduce stellar abundance patterns \citep[e.g.,][]{Qian2001_ModelAbundancesMetalpoor, Qian2007_WhereOhWhere, Kratz2007_ExplorationsRProcessesComparisons, Montes2007_NucleosynthesisEarlyGalaxy, Hansen2014_HowManyNucleosynthesis}. Such superposition of different conditions (or components) is also expected from hydrodynamical simulations, which typically produce a variety of conditions (see, e.g., Fig.~9 in KAR25).

The abundance patterns of metal-poor r-process-rich stars are ideal comparisons for r-process calculations, as these old stars formed from gas polluted by few r-process events. Their abundances are derived from observed absorption spectra using model atmospheres. Such an analysis typically leads to stellar abundances $\log_\epsilon(Z)$ with statistical uncertainties $\sigma(Z)$ (e.g., from atmospheric parameters). Additional systematic uncertainties are difficult to estimate, coming for example from missing atomic data or yet-unknown corrections from three-dimensional (3D) and non-local thermodynamical equilibrium (NLTE) effects \citep[e.g.,][]{Hansen2013_LTENonLTEThat, Lind2024_ThreeDimensionalNonlocalThermodynamic}, or internal mixing \citep[e.g.,][]{Spite2005_FirstStarsVI}.

Among the r-process-rich stars, the abundance pattern of star HD222925 \citep{Roederer2022_RprocessAllianceNearly} is almost complete and, therefore, an ideal r-process template. In contrast to the solar r-process pattern, it does not rely on theoretical s-process calculations \citep[e.g.,][]{Goriely1999_UncertaintiesSolarSystem, Prantzos2020_ChemicalEvolutionRotating}. Moreover, this star is representative of many r-process-rich stars as shown in \citet{Racca2025_RProcessAllianceExploringa} by the small scatter in the elemental abundances between the second and third peaks. In our analysis, we exclude the four elements for which only upper limits were reported in \citet{Roederer2022_RprocessAllianceNearly}: Rb ($Z=37$), Ta ($Z=73$), Bi ($Z=83$), and U ($Z=92$). For completeness, we still show those upper limits when comparing with the resulting model abundances.

We also compare to the solar r-residual abundances from \citet{Sneden2008_NeutronCaptureElementsEarly} and the abundances of the limited-r stars HD128279 \citep{Roederer2014_NewDetectionsArsenic, Roederer2022_RProcessAllianceAbundance}, and HD122563 \citep{Honda2006_NeutronCaptureElementsVery}. Although HD128279 has a lower [Ba/Sr] ratio than the typical limited-r star HD122563, it is brighter in the ultra-violet \citep{Roederer2012_NewHubbleSpace}, allowing for more elements to be derived. We assume typical uncertainties of 0.15\,dex for HD122563 and solar r.

\subsection{Statistical inference setup}
\label{sec:mcmc}

Bayes' theorem,
\begin{equation}
\label{eq:bayes}
    \prob{\params | D, I} \propto \prob{D | \params, I} \times \prob{\params | I}\,,
\end{equation}
provides the cornerstone for our statistical analysis. Here, $\params$ are the model parameters, $D$ is the available data, and $I$ encompasses all other information (such as computational methods) built into the analysis. Eq.~\eqref{eq:bayes} thus allows us to define a posterior probability distribution function (PDF) $\prob{\params | D, I}$ for $\params$ in terms of a likelihood $\prob{D | \params, I}$ and a prior $\prob{\params | I}$, both of which are themselves PDFs. For clarity, we will denote the likelihood and the prior as $\mathcal{L}(\params)$ and $\pi(\params)$, respectively.

We use an MCMC approach to sample the posterior of the parameters $\params$ of two or more components given observed stellar abundances. MCMC allows to sample a reduced subspace of parameter combinations. Another advantage is the simultaneous inference of the weights $w$, with which the different components are combined. Finally, the sampled posterior can be used for further statistical analysis, e.g., to propagate parametric uncertainties to predictions of abundances. We perform MCMC sampling using the Python package \textit{emcee} \citep{Goodman2010_EnsembleSamplersAffine, Foreman-Mackey2013_EmceeMCMCHammer}.

For each MCMC run, we fix the number of components to $N_\mathrm{comp}=2,3,\dots$ and every component is associated with four parameters, leading to a total of $4N_\mathrm{comp}$ parameters: Initial electron fraction $Y_{\mathrm{e},i}$, initial entropy $\log s_i$, expansion timescale $\log \tau_i$, and weight $\log w_i$. The length of $\params$ is thus $4N_\mathrm{comp}$. The total abundance pattern of that sample is the linear superposition of the individual patterns:
\begin{align}
\label{eq:defYSample}
    Y^\mathrm{sample}(Z)=&\sum_{i=1}^{N_\mathrm{comp}} w_i \cdot Y^\mathrm{model}_i(Y_{\mathrm{e},0},s_0,\tau;Z)\,,
\end{align}
where $Y^\mathrm{model}_i$ are the final abundance patterns (after 1~Gyr) of the closest matching ($Y_\mathrm{e},s,\tau$)-combination from KAR25. Note that the precomputed nucleosynthesis grid leads to a discrete rather than continuous parameter space of $Y_\mathrm{e}$, $s$, and $\tau$. Furthermore, the MCMC sampler treats entropy (in $\mathrm{k_B/nuc}$), expansion timescale (in s), and weights logarithmically. To simplify the notation, we drop the index $0$ in $Y_\mathrm{e,0}$ and $s_0$ from KAR25.

In order to have similar ranges of weights $w_i$ for all stars, we normalize the observed stellar abundances to Sr, by introducing a scaled $\log_\epsilon$-notation, based on the standard $\log_\epsilon$-notation:
\begin{align}
    \log_{\epsilon'}(Z)=\log_{\epsilon}(Z)-\log_{\epsilon}(Z=38)\,,
\end{align}
such that $\log_{\epsilon'}^\mathrm{obs}(Z=38)\stackrel{!}{=}0$.
Following \citet{Psaltis2024_NeutrinodrivenOutflowsElemental}, we compare the observed (logarithmic) stellar abundances, with associated statistical uncertainties $\sigma^\mathrm{obs}(Z)$, to the total abundance of a sample using the reduced chi-squared metric
\begin{align}
\label{eq:chiSquare}
    \chi^2=\frac{1}{\nu}\sum_{Z\geq Z^{\min}}\Big(\frac{\log\big(Y^\mathrm{sample}(Z)\big)-\log_{\epsilon'}^\mathrm{obs}(Z)}{\sigma^\mathrm{obs}(Z)}\Big)^2\,,
\end{align}
where we have introduced the degrees of freedom $\nu=N_\mathrm{elements}-N_\mathrm{comp}$, which depend on the number of observed elements $N_\mathrm{elements}$ for which $Z\geq Z^{\min}$. Since we are investigating the r-process contributions, we set $Z^{\min}=31$ (Ga). Via Eq.~\eqref{eq:chiSquare} we define the likelihood $\mathcal{L}(\params)$ as
\begin{align}
\label{eq:logLikelihood}
    \log \mathcal{L}(\params)=-\frac{\chi^2}{2}\,.
\end{align}
We neglect to include the normalization of the Gaussian distribution in Eq.~\eqref{eq:logLikelihood} as it will not affect the inferred posterior distributions in this case.

\begin{figure*}[t!]
    \centering
    \includegraphics[width=\linewidth, angle=0]{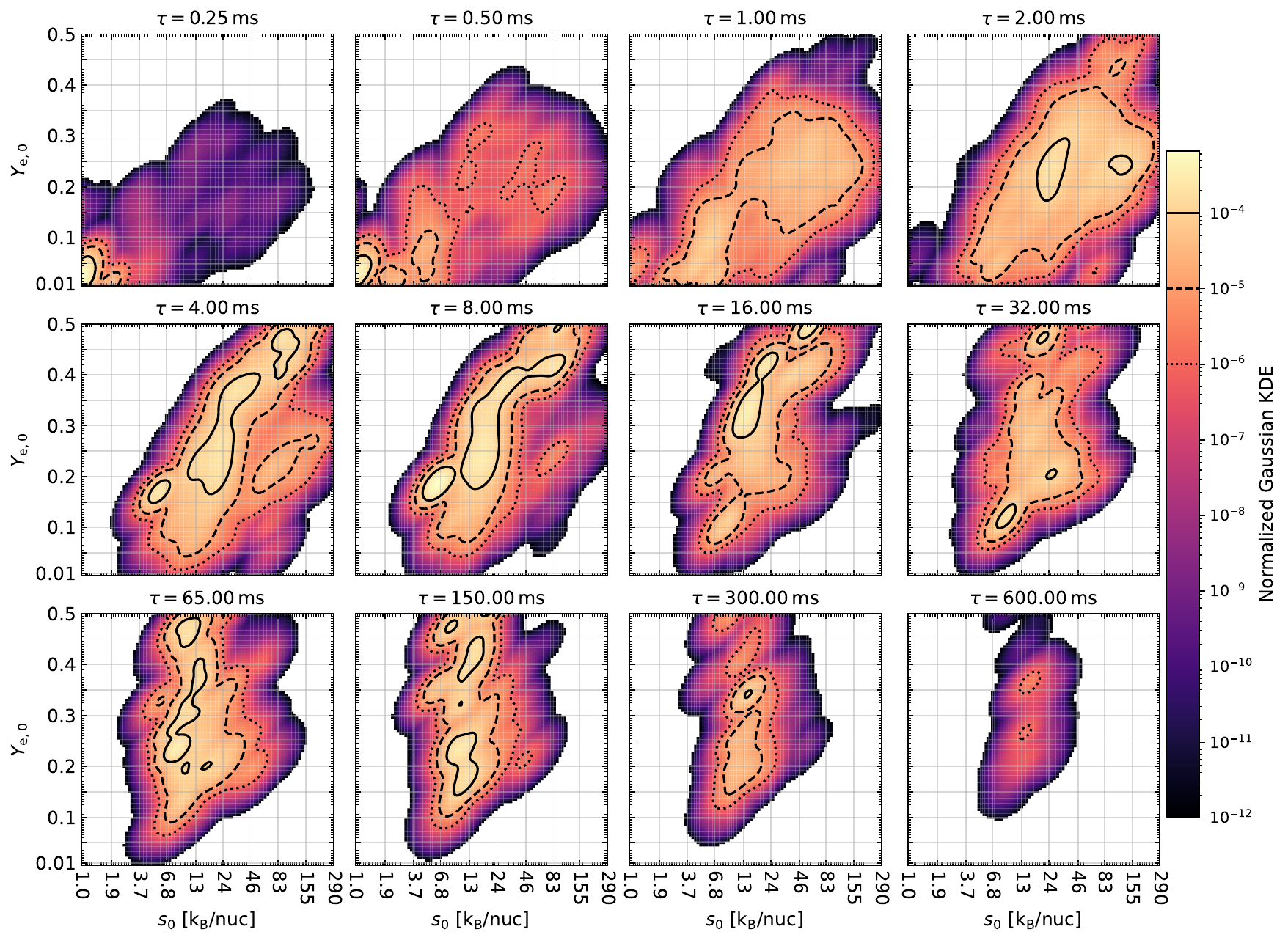}
    \caption{Normalized Gaussian KDE to quantify how often each condition is found in hydrodynamical simulations, see text for details. This quantity is used for the simulation-informed prior, defined in Eq.~(\ref{eq:logPrior}).}
    \label{fig:tracerNumberDensity}
\end{figure*}

We furthermore define the prior $\pi(\params)$, which serves multiple purposes. First, it ensures that all parameters of each component $i$ remain within their respective ranges:
\begin{align}
\label{eq:YeLimit}
    Y_{\mathrm{e},i} \in & [Y_\mathrm{e}^{\min},Y_\mathrm{e}^{\max}] = [0.005,0.5]\,,\\
\label{eq:sLimit}
    \log s_i \in & [\log s^{\min},\log s^{\max}] \\
\nonumber
               = & [\log(1)=0, \log(290)\approx2.46]\,,\\
\label{eq:tauLimit}
    \log \tau_i \in & [\log \tau^{\min},\log \tau^{\max}] \\
\nonumber
                  = & [\log(0.00025)\approx-3.60,\log(0.6)\approx-0.22]\,,\\
\label{eq:wLimit}
    \log w_i \in & [\log w^{\min},\log w^{\max}] = [1.5,5.5]\,,
\end{align}
where $s$ is in $\mathrm{k_B/nuc}$ and $\tau$ in s. If one parameter is outside its designated range, we set $\pi(\params) = 0$, causing the sample to be rejected by the MCMC sampler. Note that we only allow for parameters within the precomputed parameter grid. This is justified as our grid covers a vast set of conditions and encompasses almost all conditions found in current simulations: Our $Y_e$-range covers all neutron-rich conditions, whereas $s$ and $\tau$ cover almost all conditions found in current simulations (cf. Fig.~9 of KAR25)\footnote{Possible exceptions are extremely high entropy conditions with $s>290\,\mathrm{k_B/nuc}$, e.g., the magnetar giant flare ejecta proposed in \citet{Patel2025_DirectEvidenceRprocess, Patel2025_RProcessNucleosynthesisRadioactively}.}. Since our parameter space is limited to weak and main r-process conditions ($Y_\mathrm{e}\leq0.5$), we do not account for other nucleosynthesis processes, such as $\nu$p-, s- or i-processes. This assumption is valid as long as we limit our analysis to pure r-process stars. The limits of the weights $w$ are selected heuristically: A natural upper limit is created by the $\chi^2$ in Eq.~(\ref{eq:chiSquare}), penalizing large weights which would overproduce the observed abundances. In contrast, the lower weight limit is enforced for practical purposes to only include \enquote{relevant} conditions: Without the lower limit, any astrophysical condition could be added to an acceptable set of conditions as long as its weight is negligibly small.

Second, all components are identical, which could lead to double-counting. For example, the two parameter combinations $\{Y_{\mathrm{e},i},s_i,\tau_i,w_i,Y_{\mathrm{e},j},s_j,\tau_j,w_j\}$ and $\{Y_{\mathrm{e},j},s_j,\tau_j,w_j,Y_{\mathrm{e},i},s_i,\tau_i,w_i\}$ are identical, because both would result in the same model abundance using Eq.~(\ref{eq:defYSample}). To prevent this, we enforce the components to be ordered by increasing $Y_\mathrm{e}$. Any combination of parameters that does not fulfill $Y_{\mathrm{e},1}\leq\dots\leq Y_{\mathrm{e},{N_\mathrm{comp}}}$ will be rejected. This constraint ensures that our model satisfies identifiability.

We can be agnostic about how realistic the conditions in our parameter space are by adopting a uniform prior, i.e., one that assigns equal probability for all allowed parameters. This prior allows for astrophysical conditions that have not yet been found in hydrodynamical simulations. Alternatively, we can apply a simulation-informed prior, by using the large set of trajectories compiled in KAR25. This set includes over 29\,000 trajectories from 22 simulations by different authors \citep{Korobkin2012_AstrophysicalRobustnessNeutron, Piran2013_ElectromagneticSignalsCompact, Rosswog2013_MultimessengerPictureCompact, Bovard2017_RprocessNucleosynthesisMatter, Jacobi2023_EffectsNuclearMatter, Ricigliano2024_ImpactNuclearMatter, Fernandez2013_DelayedOutflowsBlack, Wu2016_ProductionEntireRange, Martin2015_NeutrinoDrivenWindsAftermath, Perego2014_NeutrinodrivenWindsNeutron, Bliss2020_NuclearPhysicsUncertainties, Bliss2018_SurveyAstrophysicalConditions, Obergaulinger2017_ProtomagnetarBlackHole, Reichert2021_NucleosynthesisMagnetorotationalSupernovae, Winteler2012_MagnetorotationallyDrivenSupernovae} and covers several different astrophysical scenarios, e.g., binary neutron star mergers, their disk ejecta and winds, and magneto-rotational supernovae. These trajectories were mapped to our parameter space, see Table~3 and Fig.~9 of KAR25 for more details. The simulation-informed prior is based on a Gaussian kernel density estimation (KDE)\footnote{We use the \textsc{gaussian\_kde} python method from \textsc{scipy.stats} with a bandwidth of 0.15 to ensure a satisfactory balance of smoothing and structural details.}, applied to the number density of trajectories. To guarantee an equal contribution from the three nucleosynthesis parameters ($Y_\mathrm{e}$, $s$, and $\tau$), they are each transformed to the range between 0 and 1. In addition, each trajectory is weighted by the total number of trajectories in its model. This ensures that every model is counted equally, irrespective of how many trajectories it contains. Finally, all KDE values are normalized to sum to 1 and floored at $10^{-12}$. The resulting KDE for our complete parameter space is shown in Fig.~\ref{fig:tracerNumberDensity}.

For a combination of $N_\mathrm{comp}$ components, we define the total simulation-informed prior of a sample as\footnote{Here, the mean value is favorable to the pure sum (which is typically used to combine priors), since it does not grow with $N_\mathrm{comp}$, which would otherwise decrease the relative importance of the likelihood.}:
\begin{align}
\label{eq:KDEtot}
    \mathrm{\overline{logKDE}}= \frac{1}{N_\mathrm{comp}}\sum_{i=1}^{N_\mathrm{comp}} 
    \log\Big(\mathrm{KDE}(Y_{\mathrm{e},i},s_i,\tau_i)\Big)\,.
\end{align}

Thus, we have two definitions for the prior:
\begin{align}
\label{eq:logPrior}
    \log \pi=\begin{cases}
        -\infty, \ &\text{if constraints not satisfied}\,, \\
        0, \ &\text{else (for agnostic prior)}\,, \\
        \mathrm{\overline{logKDE}}, \ &\text{else (for sim.-inf. prior)}\,,
    \end{cases}
\end{align}
where the aforementioned constraints enforce the parameters to be within the boundaries of Eqs.~(\ref{eq:YeLimit})--(\ref{eq:wLimit}) and guarantee a correct $Y_\mathrm{e}$-ordering of the components. We disregard the normalization of the prior, similarly to the likelihood.

The posterior is then defined according to Eq.~\eqref{eq:bayes} as the product of the likelihood in Eq.~(\ref{eq:logLikelihood}) and the prior in Eq.~(\ref{eq:logPrior}). It is an estimate for the total quality of a sample, measuring the agreement with both observations and (if the simulation-informed prior is used) hydrodynamical simulations.

The \textit{emcee} sampler uses multiple so-called walkers to efficiently explore the parameter space. Here, we use 64 walkers, each of which is initialized using a random draw from the uniform prior distribution spanning the parameter limits of Eqs.\,(\ref{eq:YeLimit})--(\ref{eq:wLimit}). To ensure valid initial MCMC conditions, we only allow starting conditions with finite prior values, i.e., the $Y_\mathrm{e}$-ordering has to be obeyed. To prevent the starting position from biasing the results of the MCMC sampling, the first 5\,000 steps are treated as a burn-in phase and discarded. To ensure a large enough number of uncorrelated samples, we use $10^6$ steps per walker. For all calculations, we check the MCMC convergence with the integrated auto-correlation time of all $6.4\times10^7$ samples \citep{Goodman2010_EnsembleSamplersAffine}.

\section{Results}
\label{sec:results}

\subsection{r-process star with two components}
\label{sec:2comp}

\begin{figure*}[]
    \centering
    \includegraphics[width=\linewidth]{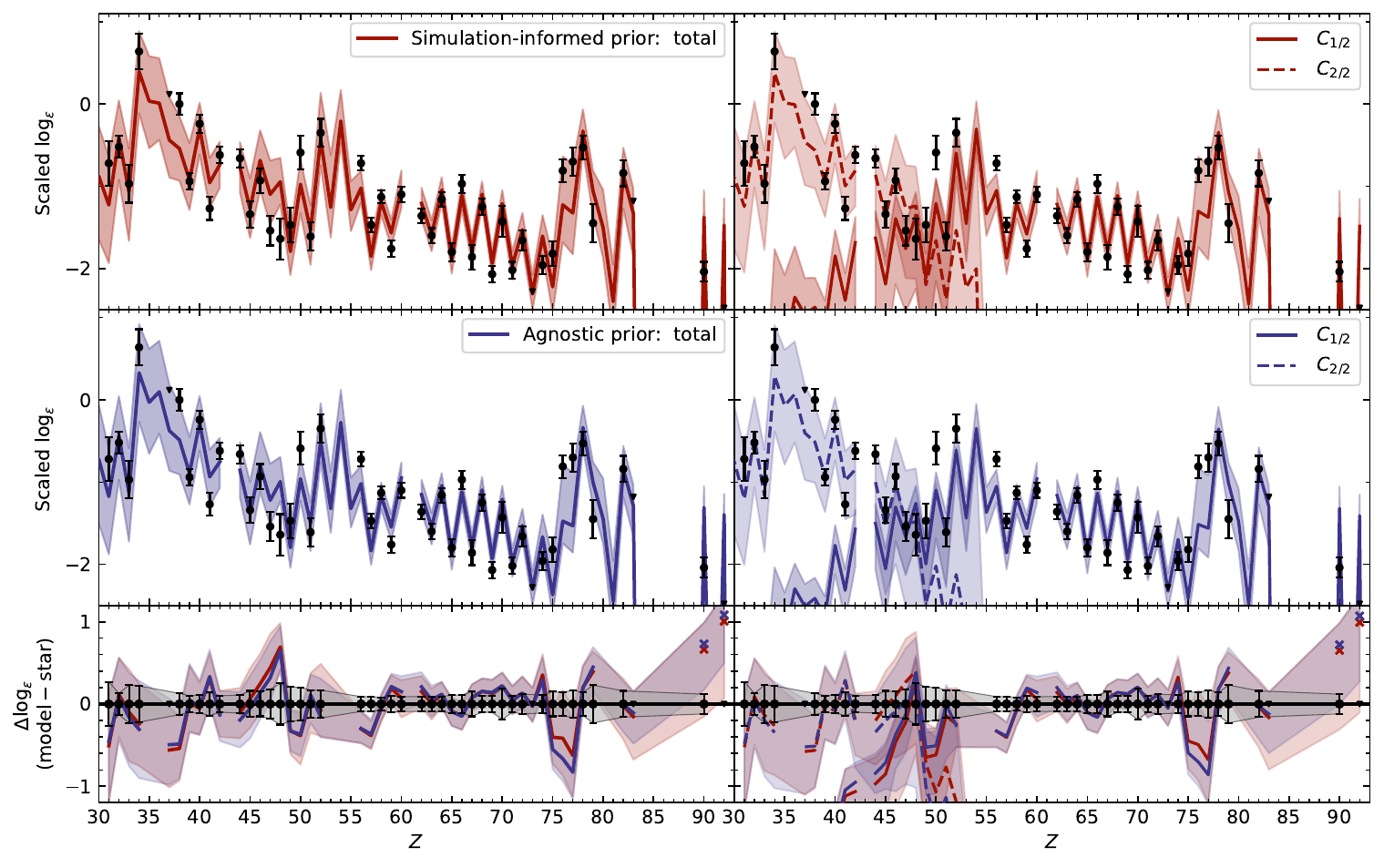}
    \caption{Resulting abundances for HD222925 with two components, corresponding to the conditions in Fig.~\ref{fig:HD222925_NP2_corner}: \textbf{Top left:} Stellar abundances from \citet{Roederer2022_RprocessAllianceNearly} with the median abundances and 68\,\% CIs for the MCMC run with the simulation-informed prior. Upper limits are shown for comparison, but not included in the fitting. \textbf{Center left:} Same for the agnostic prior. \textbf{Bottom left:} Residuals of both models and HD222925. \textbf{Right:} Same for the two individual components ($\mathrm{C}_{1/2}$ and $\mathrm{C}_{2/2}$) separately.}
    \label{fig:HD222925_NP2_abundances}
\end{figure*}

We first apply the MCMC sampler, as described in Sect.~\ref{sec:mcmc}, to the main-r star HD222925 \citep{Roederer2022_RprocessAllianceNearly} using two components and compare our results based on two different priors, cf. Eq.~(\ref{eq:logPrior}). Figure~\ref{fig:HD222925_NP2_abundances} shows the median and 68\,\% credibility intervals (CIs) for the posterior predictive distributions (PPDs) of the total abundance patterns (left panels), as well as the two components separately (right panels). The resulting median abundances for both priors reproduce well the observed pattern for HD222925. We also verified that our models do not overproduce lighter elements outside of the investigated $Z$-range, e.g., iron-group nuclei. For many elements, in particular the lanthanides, the median abundances of the models are within the observational uncertainties. Larger discrepancies (above 0.5\,dex) are found for the following elements: Ga ($Z=31$); Sr ($Z=38$); Ag and Cd ($Z=47-48$); Re, Os, and Ir ($Z=75-77$); Th ($Z=90$).

The 68\,\% CIs of the PPDs in Fig.~\ref{fig:HD222925_NP2_abundances} quantify the abundance variation of each element within the set of MCMC samples. Lanthanides ($Z=57-71$) are produced very robustly by the first component (with lower $Y_\mathrm{e}$), with a scattering of only 0.2--0.4\,dex. Several different nuclear-physics features were suggested in previous works to explain their robust production \citep[e.g.,][]{Schramm1971_SynthesisSuperheavyElements, Surman1997_SourceRareEarthElement, Goriely2013_NewFissionFragment, Eichler2015_RoleFissionNeutron, Mumpower2017_ReverseEngineeringNuclear, Sprouse2020_PropagationStatisticalUncertainties, Pallas2023_StudyDecayProperties}. In contrast, lighter elements ($31\lesssim Z\lesssim52$) in our models are produced by the second component (with higher $Y_\mathrm{e}$) and have a larger scattering of around 0.5--1\,dex. A similar variation is found in metal-poor stars \cite[see, e.g.,][]{Sneden2008_NeutronCaptureElementsEarly}. The third-peak elements Re, Os, and Ir ($75-77$) also have large variations around 1\,dex. The heaviest elements Bi, Th, and U ($Z=83$, $90$, and $92$) also show wide uncertainty bands of 1\,dex. Some of these discrepancies and variability may be an indication of lacking accuracy in the nuclear physics input. The abundances on the right side of the second peak (Ba and La; $Z=56-57$) and on the left side of the third peak (Re, Os, and Ir; $Z=75-77$) are strongly affected by nuclear masses \cite[see, e.g.,][]{Pfeiffer2001_NuclearStructureStudies, Arcones2012_NuclearCorrelationsProcess, Martin2016_ImpactNuclearMass, Mendoza-Temis2015_NuclearRobustnessProcess, Kuske2026_ProcessNucleosynthesisInitio, Li2026_ImplicationsWeakeningN126}. Observed abundances of these elements provide powerful constraints of nuclear mass models. Similarly, Th production strongly depends on nuclear masses and fission barriers \cite[see, e.g.,][]{Eichler2019_ProbingProductionActinides, Holmbeck2019_ActiniderichActinidepoorRprocessenhanced}.

\begin{figure*}[t!]
    \centering
    \includegraphics[width=0.9\linewidth]{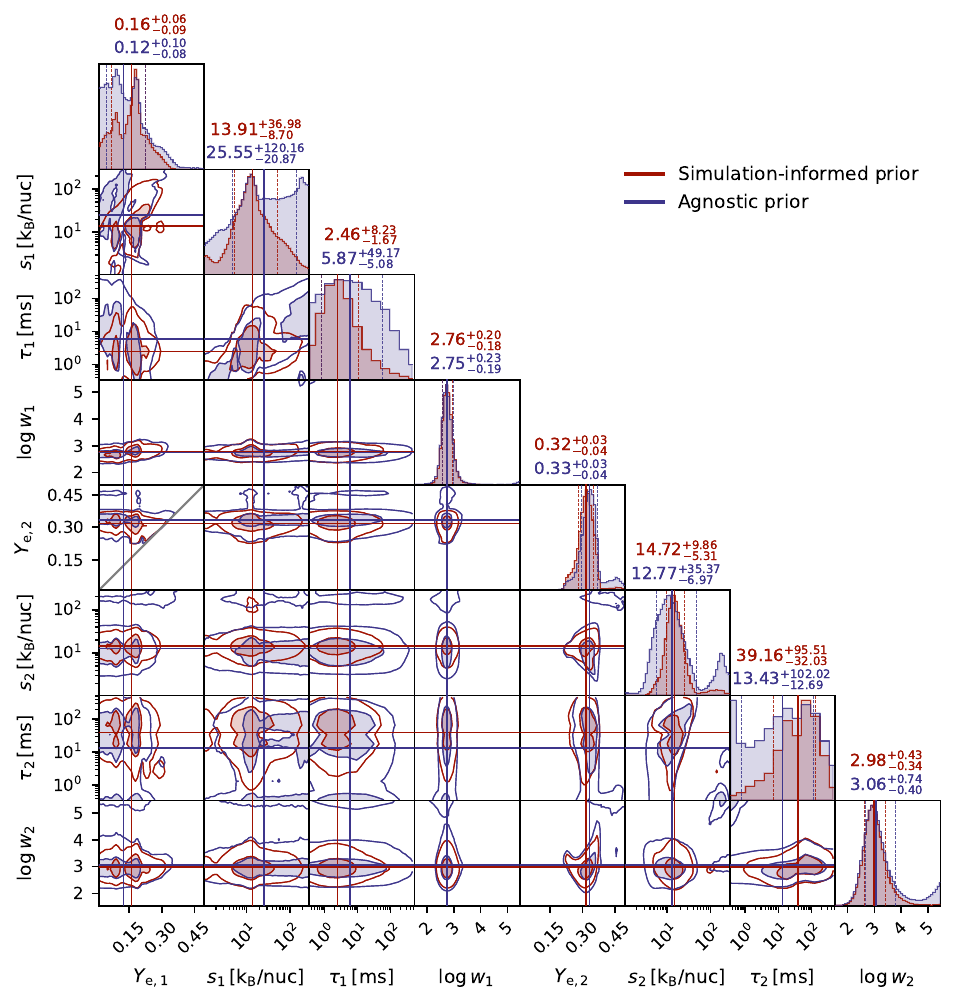}
    \caption{Corner plot of the posteriors for the abundance fit of the main-r star HD222925 using two components with both agnostic (blue) and simulation-informed (red) priors. Contour lines mark the $1\sigma$ (filled) and $2\sigma$ (transparent) bands, while vertical and horizontal solid (dashed) lines correspond to the median ($1\sigma$) values, which are also given above each column. Note that a $1\sigma$ ($2\sigma$) interval approximately corresponds to a 68\,\% (95\,\%) CI for a one-dimensional distribution and a 39\,\% (86\,\%) CI for a two-dimensional distribution. The corresponding abundances are shown in Fig.~\ref{fig:HD222925_NP2_abundances}.}
    \label{fig:HD222925_NP2_corner}
\end{figure*}

The posterior distributions for HD222925 with two components are shown in Fig.~\ref{fig:HD222925_NP2_corner} for both the agnostic and simulation-informed prior. The median and 68\,\% CIs using both priors are summarized above each column. There are some conditions excluded by the simulation-informed prior that still reproduce HD222925 abundances: very low $Y_\mathrm{e}$, very high $s$, and intermediately high $\tau$ for component 1 and high $Y_\mathrm{e}$, very high $s$, and very small $\tau$ for component 2. However, these conditions do not lead to an improvement in the resulting abundances, which are extremely similar for both priors, as shown in Fig.~\ref{fig:HD222925_NP2_abundances}.

Our results show that two components may be enough to explain the typical r-process pattern found in many old stars. This is in agreement with other works based on observations \citep[e.g.,][]{Qian2001_ModelAbundancesMetalpoor, Hansen2014_HowManyNucleosynthesis, Racca2025_RProcessAllianceExploringa}, nucleosynthesis \citep[e.g.,][]{Qian2007_WhereOhWhere}, and galactic chemical evolution \citep[e.g.,][]{Molero2026_ConstrainingRprocessNucleosynthesis}. These two components are sometimes called L- and H-component for the light (below second peak) and heavy r-process (second peak and beyond), respectively. However, some abundances are still not well reproduced with these two components. In the following, we will analyze whether these differences come from the limited number of components, observational data/accuracy, or other factors.

\subsection{r-process star with more components}
\label{sec:mainRComparingComponentNumbers}

\begin{figure*}[t!]
    \centering
    \includegraphics[width=1\linewidth]{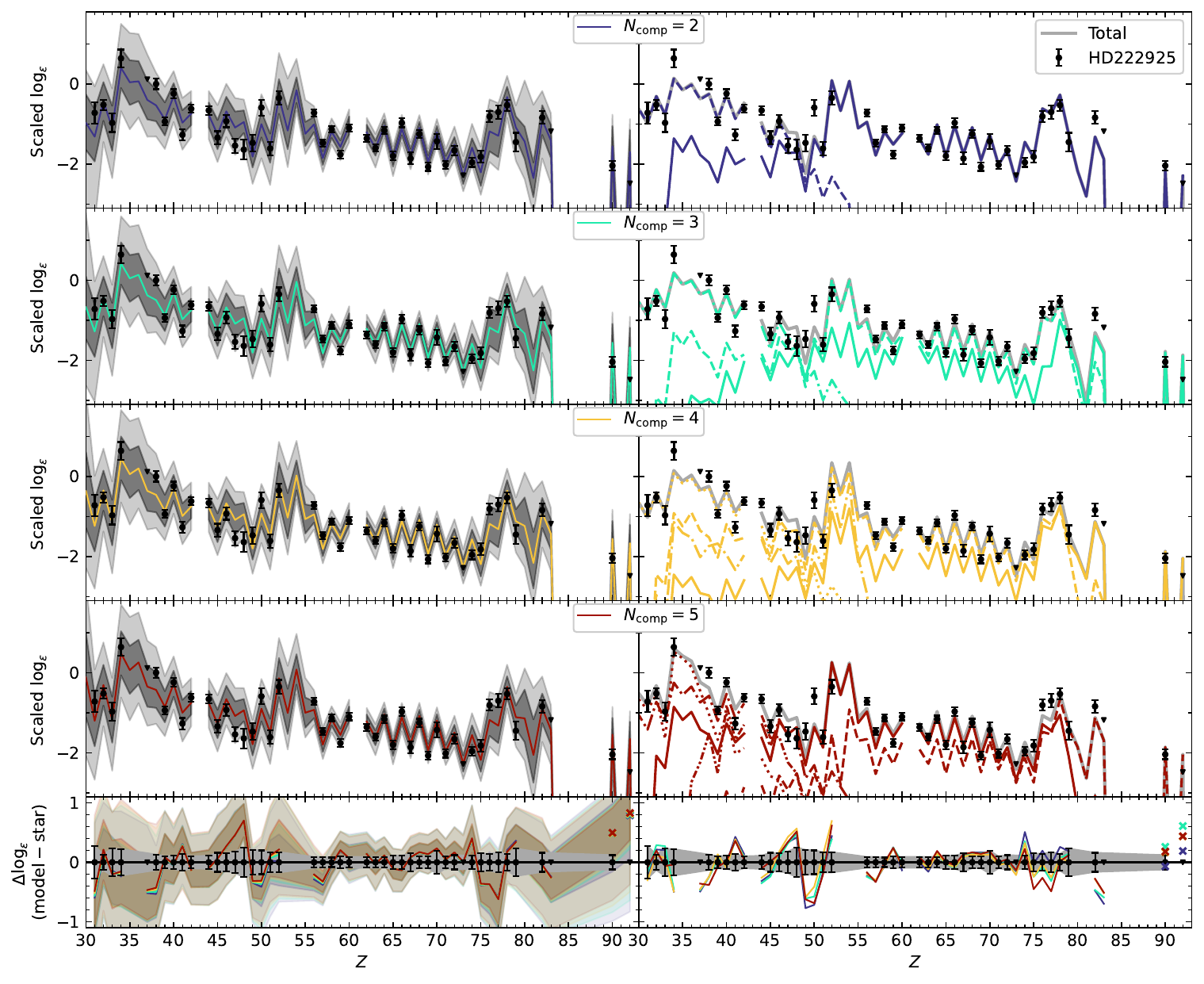}
    \caption{Abundance fits for HD222925 using different numbers of components $N_\mathrm{comp}$. \textbf{Left:} Median, 68\,\% CIs, and 95\,\% CIs of the PPDs. \textbf{Right:} Best fitting combinations with different linestyles for each component, following the legend of Fig.~\ref{fig:HD222925_NP2,3,4,5_Ye,s,tau}. \textbf{Bottom:} Residuals of models and HD222925.}
    \label{fig:HD222925_NP2,3,4,5_abundances}
\end{figure*}

We explore here the superpositions of three, four, and five components to fit HD222925. Since the two different priors result in practically identical abundances, we show only results with the simulation-informed prior here. This prior results in more realistic conditions and its smaller effective parameter space leads to a faster MCMC convergence. For each component number, the median as well as the 68\,\% and 95\,\% CIs of the predicted abundances are shown in the left panels of Fig.~\ref{fig:HD222925_NP2,3,4,5_abundances}.

Intriguingly, we find that the median values and shown CIs of the HD222925 fits are almost identical for different component numbers ($N_\mathrm{comp}=2-5$). This can be explained by investigating the corresponding best fitting combinations. These are found with a dedicated optimization algorithm to maximize the posterior, starting from the 50 best MCMC samples with 10 random perturbations each. The resulting abundances of each component are shown in the right panels of Fig.~\ref{fig:HD222925_NP2,3,4,5_abundances} and the corresponding parameters are summarized in Table~\ref{tab:bestFittingComponentsHD222925}. For $N_\mathrm{comp}=2$, the first (low-$Y_\mathrm{e}$) component produces elements from the second to the third peaks (H-component), while the second (high-$Y_\mathrm{e}$) component produces elements from the first to the second peaks (L-component). Most of the additional components that are added for $N_\mathrm{comp}=3-5$ result in improved abundance agreements for the $Z$-range of either the H- or the L-component. For example, the best fits for $N_\mathrm{comp}=3$ and $4$ each consist of two H-components and one or two L-components, respectively. This leads to minor improvements in the agreement of the best fits. However, crucially for the behavior of the median abundances, the additional components that are added for $N_\mathrm{comp}=3-5$ come largely from the same set of L- and H-conditions as for $N_\mathrm{comp}=2$. Therefore, a larger number of components effectively only increases the number of components that are averaged, without changing the resulting total abundances. This result suggests that two components are enough to reproduce most elements of the abundance pattern found in HD222925 and similar stars.

\begin{table}[]
    \centering
    \caption{Parameters of best fitting combinations for HD222925 using different component numbers $N_\mathrm{comp}$, corresponding to the right panels of Fig.~\ref{fig:HD222925_NP2,3,4,5_abundances}.}
    \begin{tabular}{cccccc}
        \hline
        \hline
        $C_{i/N_\mathrm{comp}}$ & $Y_{\mathrm{e},i}$ & $s_i$ & $\tau_i$ & $\log w_i$ \\
        & & [$\mathrm{k_B/nuc}$] & [ms] & \\
        \hline
        $C_{1/2}$ & 0.170 & 13.80 &  8 & 2.86 \\
        $C_{2/2}$ & 0.330 & 15.50 & 16 & 2.81 \\
        \hline
        $C_{1/3}$ & 0.050 &  1.05 &  0.25 & 2.07 \\
        $C_{2/3}$ & 0.185 & 16.50 &  8    & 2.71 \\
        $C_{3/3}$ & 0.330 & 13.80 & 16    & 2.83 \\
        \hline
        $C_{1/4}$ & 0.150 & 15.50 &  4  & 2.21 \\
        $C_{2/4}$ & 0.190 & 13.80 &  8  & 2.83 \\
        $C_{3/4}$ & 0.205 & 11.50 & 150 & 2.25 \\
        $C_{4/4}$ & 0.335 & 15.50 & 16  & 2.76 \\
        \hline
        $C_{1/5}$ & 0.210 & 17.50 &  4 & 2.84 \\
        $C_{2/5}$ & 0.255 & 70.00 &  2 & 2.26 \\
        $C_{3/5}$ & 0.330 & 17.50 & 16 & 2.46 \\
        $C_{4/5}$ & 0.355 & 12.20 & 65 & 3.05 \\
        $C_{5/5}$ & 0.380 & 41.00 &  8 & 1.63 \\
        \hline
    \end{tabular}
    \label{tab:bestFittingComponentsHD222925}
\end{table}

\begin{figure*}
    \centering
    \includegraphics[width=\linewidth]{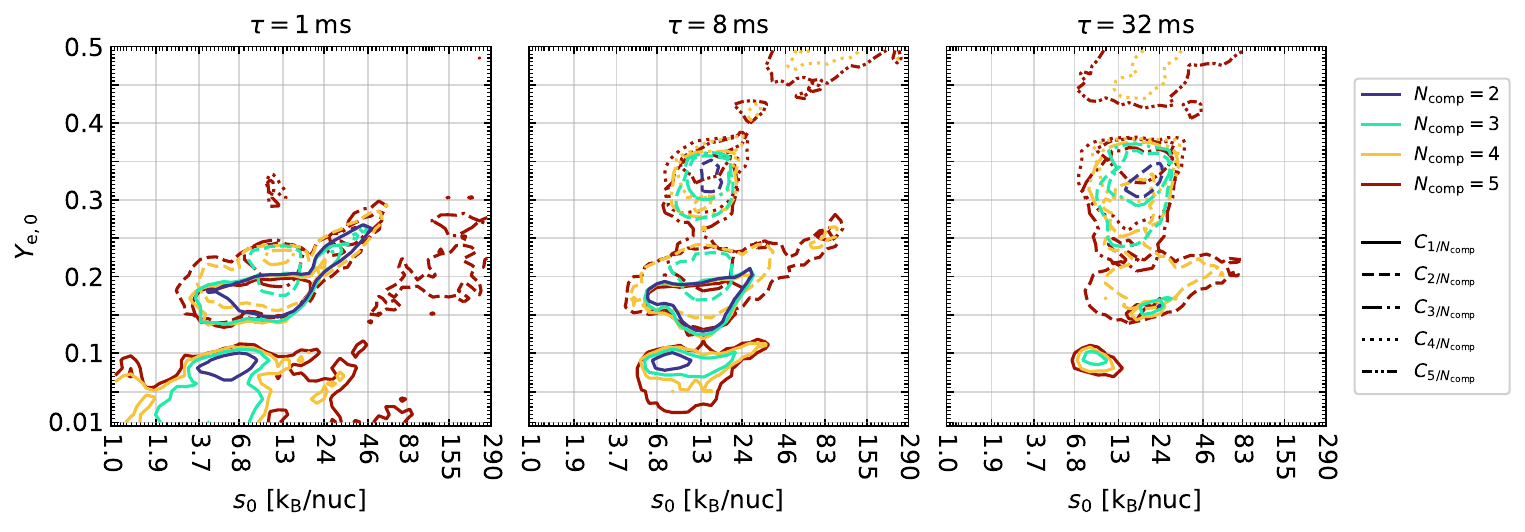}
    \caption{Parameter space regions (for three typical expansion timescales) within which 30\,\% of (unweighted) samples are found that produce the abundances of HD222925 for $N_\mathrm{comp}=2-5$, corresponding to the abundances in Fig.~\ref{fig:HD222925_NP2,3,4,5_abundances}.}
    \label{fig:HD222925_NP2,3,4,5_Ye,s,tau}
    \vspace{0.5cm}
\end{figure*}

The parameter space regions that include 30\,\% of all (unweighted) samples are shown in Fig.~\ref{fig:HD222925_NP2,3,4,5_Ye,s,tau} for each component, focusing on three typical expansion timescales. Generally, higher $N_\mathrm{comp}$ allow for a more diverse range of conditions, leading to more spread-out 30\,\% regions. For the first component these regions are centered around $Y_\mathrm{e}\approx0.06-0.2$, $s\approx5-30\,\mathrm{k_B/nuc}$, and $\tau\approx0.5-16\,\mathrm{ms}$, while for the second component they evolve around $Y_\mathrm{e}\approx0.30-0.35$, $s\approx10-20\,\mathrm{k_B/nuc}$, and $\tau\approx8-150\,\mathrm{ms}$. For $N_\mathrm{comp}=4$ and $5$, conditions with $Y_\mathrm{e}\gtrsim0.4$ and $s=10-50\,\mathrm{k_B/nuc}$ become common. These conditions produce only elements up to the first r-process peak and correspond to nucleosynthesis group G1 in KAR25.

For most elements, the abundances of the best fits are similar to the median abundances, cf. lower panels in Fig.~\ref{fig:HD222925_NP2,3,4,5_abundances}. Some of the heaviest elements, i.e., Os ($Z=76$), Ir ($Z=77$), Au ($Z=79$), and Th ($Z=90$), are improved in the best fits compared to the median, because they are particularly sensitive to the astrophysical conditions. In contrast, for other elements the median abundances result in better agreement than the best single fit, e.g., In ($Z=49$), Sn ($Z=50$), Te ($Z=52$), and Pb ($Z=82$). Except for the latter, they are all in the transition region between the L- and H-components and have contributions from both. These elements benefit from the superposition of many conditions ($\gg5$) that enter the median -- some of which overproduce and some of which underproduce them.

A similar fitting of the HD222925 abundances was performed by \citet{Holmbeck2023_HD222925New}, also using similar nuclear inputs. In contrast to our likelihood, which includes all elements, see Eqs.~(\ref{eq:chiSquare})--(\ref{eq:logLikelihood}), their matching was based on few selected elemental ratios. In addition, their framework combined a fixed number of 15 components from a set of two trajectories (i.e., two $s$- and $\tau$-combinations) with varied $Y_\mathrm{e}$. Our approach here significantly expands the parameter space of $s$ and $\tau$ and allows us to investigate the number of required components. For $Z\gtrsim38$ our results agree well with those of \citet{Holmbeck2023_HD222925New}. In particular, we confirm the gap of conditions with $Y_\mathrm{e}\approx0.12$ (see Fig.~\ref{fig:HD222925_NP2,3,4,5_Ye,s,tau}), due to unfavorable fission contributions and we find abundance differences for similar elements. However, for lighter elements, in particular As and Se ($Z=33-34$), our results are significantly (almost 2\,dex) closer to the observed abundances of HD222925. This is likely due to the lack of conditions with $Y_\mathrm{e}\approx0.33$ and $s\approx12-15\,\mathrm{k_B/nuc}$ in \citet{Holmbeck2023_HD222925New}, which we find to be dominant for the production of these elements in HD222925.

In summary, the abundances of most elements in HD222925 can be reproduced using just two r-process components: An L-component for elements between the first and second peaks and an H-component for elements at the second peak and heavier. However, some elements in between ($49\lesssim Z\lesssim52$) have contributions from both components and require more components. Since a wide range of neutron-rich astrophysical conditions are covered here, the remaining differences are mainly due to observational and nuclear uncertainties.

\begin{figure*}[ht!]
    \centering
    \includegraphics[width=1\linewidth]{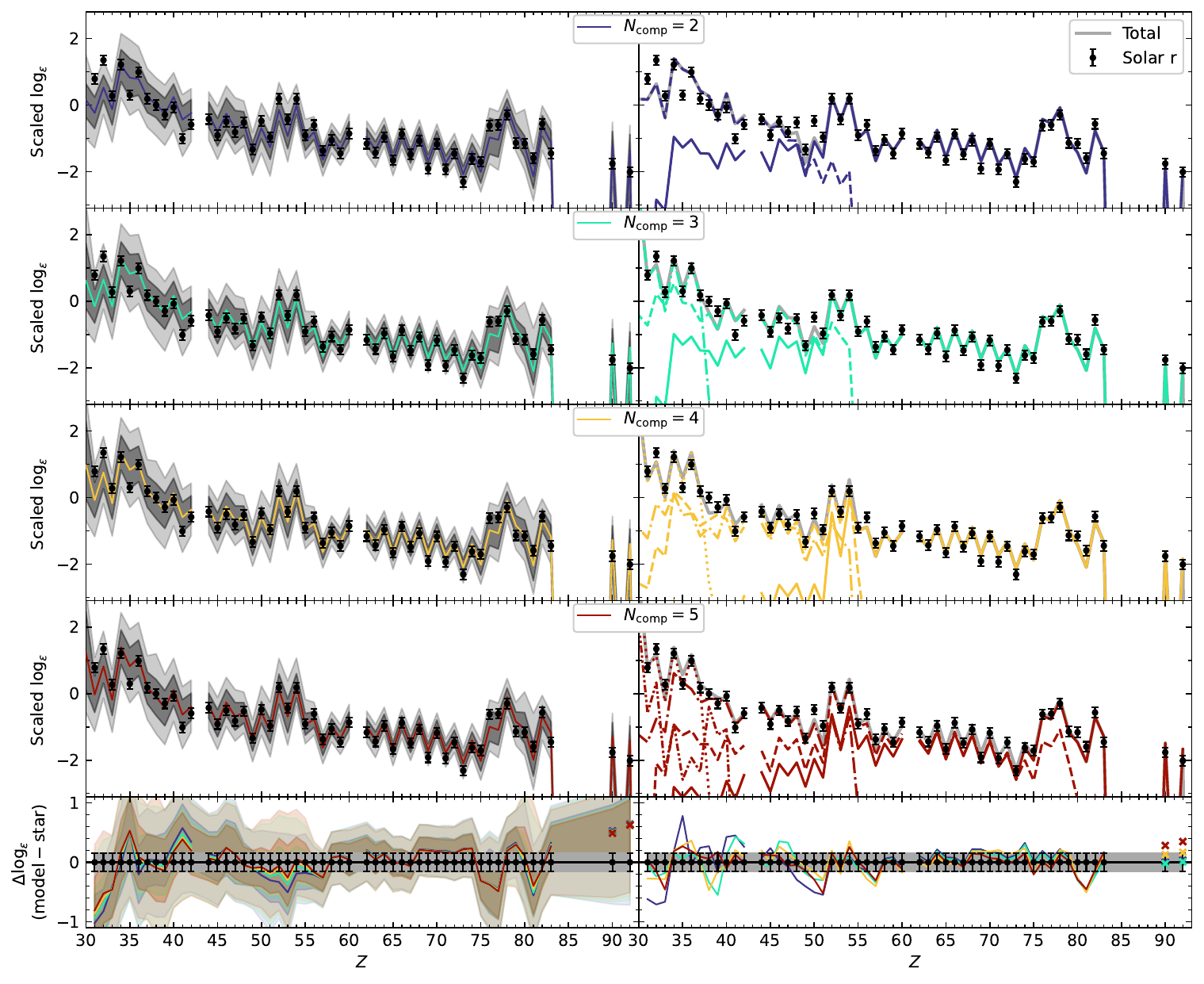}
    \caption{Same as Fig.~\ref{fig:HD222925_NP2,3,4,5_abundances} for the solar r-process residuals. The corresponding best fitting conditions are given in Table~\ref{tab:bestFittingComponentsSolarR}. Note that at least three components are necessary to reproduce the lighter elements around $Z\approx30-35$.}
    \label{fig:solarR_NP2,3,4,5_abundances}
\end{figure*}

\subsection{r-process star vs. solar r-process}
\label{sec:comp2solar}

In order to understand the discrepancies between observed and fitted abundances, it is useful to also compare the differences between solar r and HD222925. We now apply the MCMC fitting with the simulation-informed prior and two components to the solar r-residual abundances from \citet{Sneden2008_NeutronCaptureElementsEarly}. The solar r-abundances of the lighter elements Ga, Ge, and As ($Z=31-33$) are around 1--2\,dex higher than in HD222925. This indicates significant contributions from additional nucleosynthesis processes in the Sun, cf. the discussion in Section~\ref{sec:intro}. Alternatively, it could be caused by problems with the observations and/or extraction of the solar r abundances from calculated solar s-process abundances. The lower metallicity  of HD222925 points to few or no additional nucleosynthesis processes besides the r-process \citep{Roederer2022_RprocessAllianceNearly}. Moreover, the good agreement of our two-component r-process models for Ge and As favors a dominant r-process contribution for HD222925. The discrepancy for Ga could potentially originate from a yet unidentified Fe line, in which case the reported Ga abundance should be interpreted as an upper limit \citep{Roederer2022_RprocessAllianceNearly} and would thus agree with our models.

As shown in Fig.~\ref{fig:solarR_NP2,3,4,5_abundances}, for solar r the lighter neutron-capture elements ($31\lesssim Z\lesssim35$) cannot be reproduced with $N_\mathrm{comp}=2$. We find that a third component reproduces the first peak well with conditions of $Y_\mathrm{e}\approx0.42$, $s\approx12-15\,\mathrm{k_B/nuc}$, and $\tau\approx150\,\mathrm{ms}$, cf. Table~\ref{tab:bestFittingComponentsSolarR}. The median abundances of solar r improve with increased $N_\mathrm{comp}$ for certain elements: Ga, Ge, and As ($Z=31-33$), Nb ($Z=41$), and the elements in the transition regime of the L- and H-components ($50\lesssim Z\lesssim53$). Reproducing the shape of Sr, Y, and Zr ($Z=38-40$) requires at least $N_\mathrm{comp}=5$.

\begin{table}[t!]
    \centering
    \caption{Same as Table~\ref{tab:bestFittingComponentsHD222925} for solar r, corresponding to the right panels of Fig.~\ref{fig:solarR_NP2,3,4,5_abundances}.}
    \begin{tabular}{cccccc}
        \hline
        \hline
        $C_{i/N_\mathrm{comp}}$ & $Y_{\mathrm{e},i}$ & $s_i$ & $\tau_i$ & $\log w_i$ \\
        & & [$\mathrm{k_B/nuc}$] & [ms] & \\
        \hline
        1/2 & 0.170 & 12.20 &   2 & 3.08 \\
        2/2 & 0.320 & 10.80 &  65 & 3.80 \\
        \hline
        1/3 & 0.170 & 12.20 &   2 & 3.05 \\
        2/3 & 0.290 & 12.20 &  16 & 3.06 \\
        3/3 & 0.420 & 14.50 & 150 & 5.33 \\
        \hline
        1/4 & 0.225 & 35.00 &   2 & 2.86 \\
        2/4 & 0.230 & 10.80 & 150 & 3.02 \\
        3/4 & 0.285 & 13.80 &  16 & 2.67 \\
        4/4 & 0.415 & 13.00 &  65 & 4.79 \\
        \hline
        1/5 & 0.160 &  6.00 &   1 & 2.67 \\
        2/5 & 0.215 & 17.50 &   2 & 2.87 \\
        3/5 & 0.290 &  8.25 & 150 & 3.06 \\
        4/5 & 0.420 & 12.20 & 150 & 5.47 \\
        5/5 & 0.450 & 17.50 & 150 & 5.21 \\
        \hline
    \end{tabular}
    \label{tab:bestFittingComponentsSolarR}
\end{table}

The Sr/Y ratio derived for HD222925 (and many other metal-poor stars, e.g., \citealt{Sneden2008_NeutronCaptureElementsEarly, Racca2025_RProcessAllianceExploringa}) is significantly larger than in solar r and our models, cf. Fig.~\ref{fig:HD222925_NP2,3,4,5_abundances}. The recommended Sr abundance of HD222925 is that of singly ionized Sr~II \citep{Roederer2018_RProcessAllianceComprehensive}, which has been shown to have smaller NLTE corrections than Sr~I \citep{Hansen2013_LTENonLTEThat}. Intriguingly, \citet{Roederer2018_RProcessAllianceComprehensive} also report a Sr abundance of Sr~I, which is 0.41\,dex higher than that from Sr~II and would agree well with our models. Another possible solution to the Sr/Y discrepancy is an NLTE correction for Y, which could increase the Y abundance by around 0.5\,dex, resulting in agreement with solar r \citep{Storm2023_ObservationalConstraintsOrigin}. Furthermore, the solar r-process abundances of these s-process peak elements are subject to particularly large uncertainties \citep[see, e.g.,][]{Goriely1999_UncertaintiesSolarSystem}.

The low HD222925 abundances of Pd, Ag, and Cd ($Z=46-48$) are not reproduced by our models. Instead, the models result in abundances similar to solar r. Note that the solar Cd abundance is based on meteoritic data, because the photospheric lines are heavily blended \citep{Asplund2009_ChemicalCompositionSun}. A correlation between the Cd abundances of several stars and their stellar parameters was recently found by \citet{Shah2024_RProcessAllianceDetailed}, indicating strong yet unknown NLTE effects. Similarly, observational uncertainties could be a possible explanation for the abundance discrepancies in W ($Z=74$). Due to its weak absorption lines and lacking atomic data \citep{Roederer2022_RprocessAllianceNearly}, abundance derivations of W have only been performed in few stars \cite[see, e.g.,][]{Roriz2024_TungstenBariumStars}.

\subsection{Additional stellar abundance patterns}
\label{sec:comparingStars}

\begin{figure*}[t!]
    \centering
    \includegraphics[width=1\linewidth]{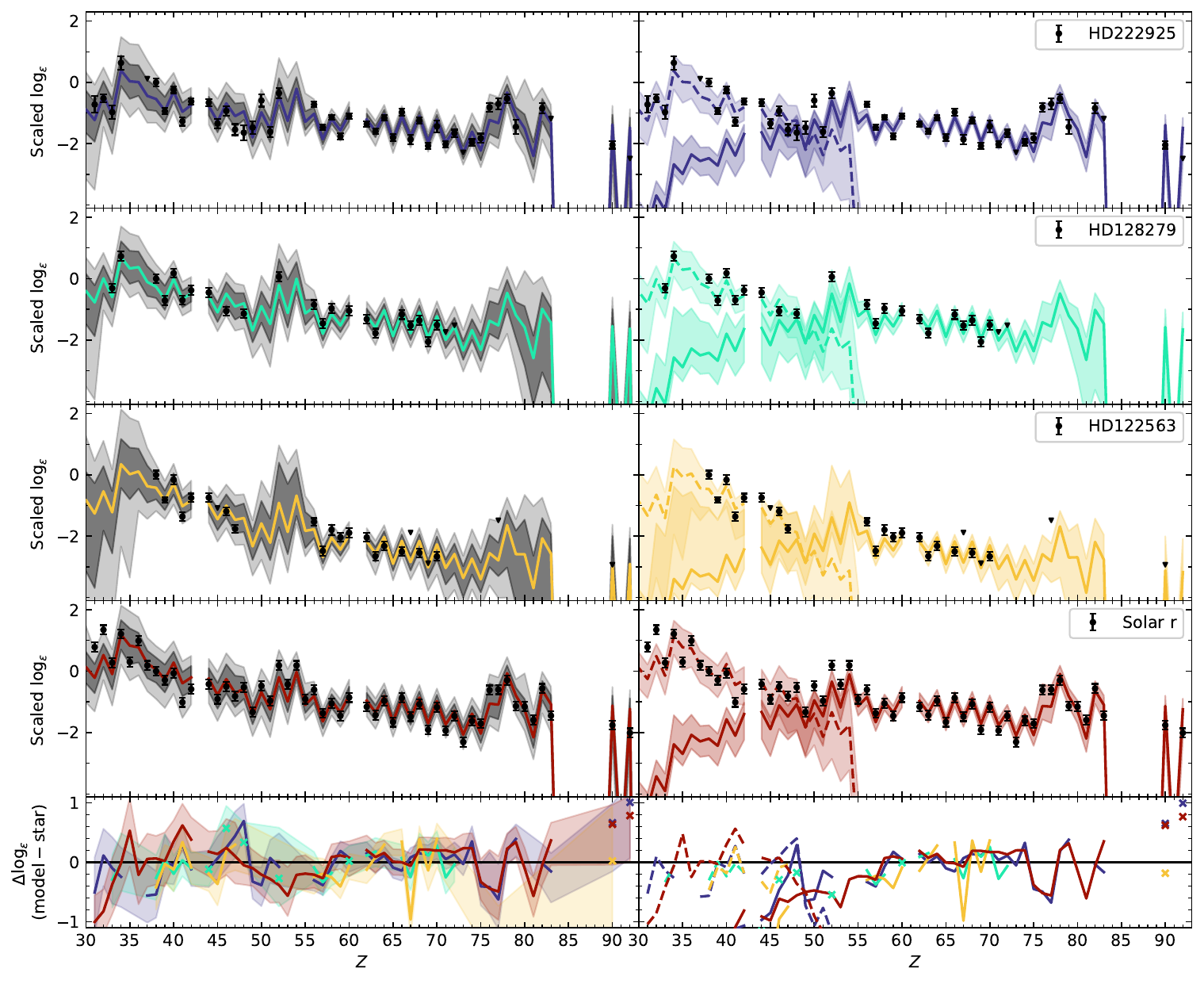}
    \caption{Same as Fig.~\ref{fig:HD222925_NP2_abundances} but for different stellar abundances, using the simulation-informed prior: The main-r star HD222925 \citep{Roederer2022_RprocessAllianceNearly}, the slightly limited-r star HD128279 \citep{Roederer2014_NewDetectionsArsenic, Roederer2022_RProcessAllianceAbundance}, the typical limited-r star HD122563 \citep{Honda2006_NeutronCaptureElementsVery}, as well as the solar r-process residuals \citep{Sneden2008_NeutronCaptureElementsEarly}. The total abundances in the left panels show the medians, $1\sigma$ and $2\sigma$ bands, whereas for visual clarity the $2\sigma$ bands are left out in the residual plot. Likewise, the component abundances on the right include the medians and $1\sigma$ bands, whereas the corresponding residuals are only shown for the medians. The corresponding conditions are shown in Fig.~\ref{fig:HD222925_HD128279_solarR_NP2_corner} and summarized in Table~\ref{tab:starsFittingComponents}.}
    \label{fig:HD222925_HD128279_solarR_NP2_abundances}
\end{figure*}

Next, we extend our study to stars with a low enrichment of heavy r-process elements (H-component), namely stars with high Sr/Ba ratios. These are also known as r-limited or Honda-like stars and a typical example is HD122563 \citep{Honda2006_NeutronCaptureElementsVery} with $\mathrm{[Sr/Ba]}=0.78$. We also include the star HD128279 \citep{Roederer2014_NewDetectionsArsenic, Roederer2022_RProcessAllianceAbundance} in this category, despite its smaller Sr/Ba ratio of $\mathrm{[Sr/Ba]}=0.10$. As seen in Fig.~\ref{fig:HD222925_HD128279_solarR_NP2_abundances}, two components result in excellent abundance agreement for most observed elements in HD128279 and HD122563. No significant improvement is found when using more components.

\begin{figure*}[t!]
    \centering
    \includegraphics[width=0.9\linewidth]{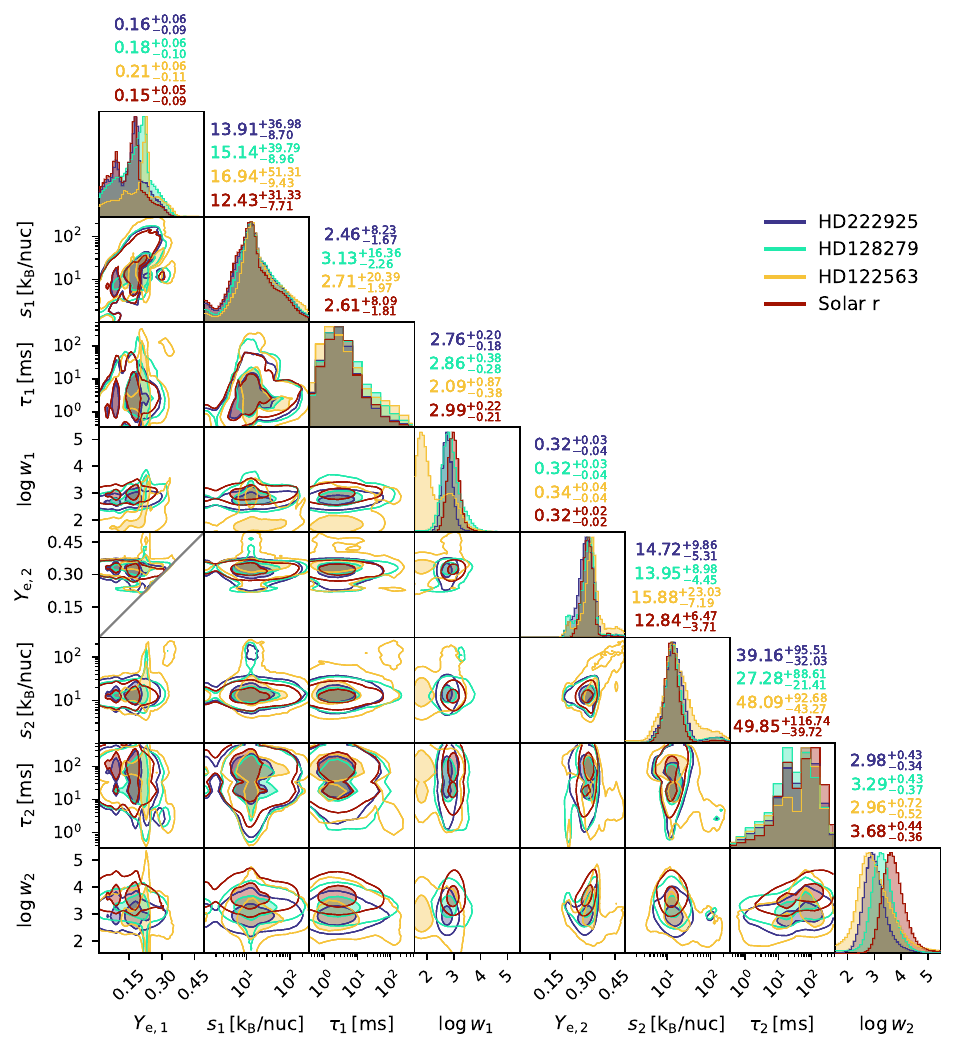}
    \caption{Same as Fig.~\ref{fig:HD222925_NP2_corner} but for different stars, using the simulation-informed prior. The color scheme is identical to Fig.~\ref{fig:HD222925_HD128279_solarR_NP2_abundances}, where the corresponding abundances are shown. Significant differences in the parameters are limited to the relative weights of the two components.}
    \label{fig:HD222925_HD128279_solarR_NP2_corner}
\end{figure*}

\begin{table*}[]
    \centering
    \caption{Median and 68\,\% CIs of all parameters for the two component fits of the four investigated stars, corresponding to the abundances in Fig.~\ref{fig:HD222925_HD128279_solarR_NP2_abundances} and the parameters in Fig.~\ref{fig:HD222925_HD128279_solarR_NP2_corner}.}
    \begin{tabular}{l|cccccc}
        \hline
        \hline
        Star & $C_{i/N_\mathrm{comp}}$ & $Y_{\mathrm{e},i}$ & $s_i$ & $\tau_i$ & $\log w_i$ \\
        & & & [$\mathrm{k_B/nuc}$] & [ms] & \\
        \hline
        HD222925 & $C_{1/2}$ & $0.16^{+0.06}_{-0.09}$ & $13.91^{+36.98}_{-8.70}$ & $2.46^{+8.23}_{-1.67}$     & $2.76^{+0.20}_{-0.18}$ \\
                 & $C_{2/2}$ & $0.32^{+0.03}_{-0.04}$ & $14.72^{+9.86}_{-5.31}$  & $39.16^{+95.51}_{-32.03}$  & $2.98^{+0.43}_{-0.34}$ \\
        \hline
        HD128279 & $C_{1/2}$ & $0.18^{+0.06}_{-0.10}$ & $15.14^{+39.79}_{-8.96}$ & $3.13^{+16.36}_{-2.26}$    & $2.86^{+0.38}_{-0.28}$ \\
                 & $C_{2/2}$ & $0.32^{+0.03}_{-0.04}$ & $13.95^{+8.98}_{-4.45}$  & $27.28^{+88.61}_{-21.41}$  & $3.29^{+0.43}_{-0.37}$ \\
        \hline
        HD122563 & $C_{1/2}$ & $0.21^{+0.06}_{-0.11}$ & $16.94^{+51.31}_{-9.43}$ & $2.71^{+20.39}_{-1.97}$    & $2.09^{+0.87}_{-0.38}$ \\
                 & $C_{2/2}$ & $0.34^{+0.04}_{-0.04}$ & $15.88^{+23.03}_{-7.19}$ & $48.09^{+92.68}_{-43.27}$  & $2.96^{+0.72}_{-0.52}$ \\
        \hline
        Solar r  & $C_{1/2}$ & $0.15^{+0.05}_{-0.09}$ & $12.43^{+31.33}_{-7.71}$ & $2.61^{+8.09}_{-1.81}$     & $2.99^{+0.22}_{-0.21}$ \\
                 & $C_{2/2}$ & $0.32^{+0.02}_{-0.02}$ & $12.84^{+6.47}_{-3.71}$  & $49.85^{+116.74}_{-39.72}$ & $3.68^{+0.44}_{-0.36}$ \\
        \hline
    \end{tabular}
    \label{tab:starsFittingComponents}
\end{table*}

Although the four abundance patterns investigated in Fig.~\ref{fig:HD222925_HD128279_solarR_NP2_abundances} show significant differences, we find that the two components that best produce them have similar conditions, see Fig.~\ref{fig:HD222925_HD128279_solarR_NP2_corner} and Table~\ref{tab:starsFittingComponents}. Even the $1\sigma$ and $2\sigma$ regions of the four stars overlap in the corner plot for most parameters. An exception to this is the electron fraction of the first component, which tends to be higher for the limited-r stars than for HD222925 and solar r. Significant differences between the stars are restricted to the weights, in particular the relative scaling of the two components: We find for HD222925 $\bar w_2/\bar w_1\approx1.7$, for HD128279 $\bar w_2/\bar w_1\approx2.7$, for HD122563 $\bar w_2/\bar w_1\approx7.4$, and for solar r $\bar w_2/\bar w_1\approx4.9$. These results indicate that the same two sets of astrophysical conditions (with slight variation in the electron fraction of the first component) are the dominant contributions to the r-process elements in all four considered stars (although solar r requires additional conditions, as discussed in Section~\ref{sec:comp2solar}), but with different relative scaling.

A similar \enquote{universality} was found by \citet{Roederer2022_RProcessAllianceAbundance} and \citet{Racca2025_RProcessAllianceExploringa} for sets of eight and ten metal-poor r-process-rich stars, respectively. For most elements, they found similar abundances (within about 0.13\,dex) between all stars when separately normalizing the lighter elements ($34\leq Z\leq52$) to Zr and the heavier ones ($56\leq Z\leq79$) to Eu. Larger dispersions were found for elements Ru, Rh, Pd, Ag, Cd, and Sn ($44\leq Z\leq50$), which were interpreted as possible signs of fission fragments by \citet{Roederer2023_ElementAbundancePatterns}. Alternative explanations discussed above are: Pd, Ag, and Cd could have strong NLTE effects and In and Sn are in the transition regime of the two components and require superpositions of many components to be produced. Differences in the superpositions could, therefore, also explain the abundance variations. Larger, homogeneously analyzed observational studies and reduced uncertainties are required to distinguish between these explanations.

\section{Conclusions}
\label{sec:conclusions}
We have presented a Bayesian framework, based on MCMC sampling, to infer the astrophysical conditions of the r-process by combining calculated and observed abundances. Each fit combines nucleosynthesis calculations from our site-independent survey (KAR25), parameterized by the initial electron fraction, entropy, and expansion timescale, with MCMC-determined weights. Applying this approach to the r-process template star HD222925, two limited-r stars, and the solar r-process residuals, we reach the following main conclusions:

\begin{enumerate}
    \item We have tested agnostic and simulation-informed priors and both yield nearly identical abundances. This indicates that the conditions required to reproduce HD222925 are already commonly realized in current hydrodynamical simulations and that the abundances themselves do not demand exotic conditions outside the simulated parameter space.
    \item Two components are sufficient to reproduce most of the observed r-process pattern in HD222925: a low-$Y_\mathrm{e}$ component (H) that produces heavier elements from the second peak and beyond, and a higher-$Y_\mathrm{e}$ component (L) that produces lighter elements from the first to the second peak. Increasing the number of components to three, four, or five does not significantly change the median abundances, because the additional components are similar to those of the two-component case. However, these additional components allow for slightly larger variations of the astrophysical conditions, as is also expected from any natural r-process event.
    \item The same two sets of conditions dominate across stars with very different abundance patterns: the main-r star HD222925, the limited-r stars HD128279 and HD122563, and the solar r-process. The patterns differ mainly in the relative weighting of the two components. This points to a form of universality in the underlying conditions, whether these arise from different ejecta components of a single astrophysical event or from two distinct r-process sites.
    \item For solar r, additional contributions to the lighter elements around $Z\approx30-35$ are necessary to explain the observed pattern.
    \item Some abundances are not well reproduced, even with additional components. These discrepancies are concentrated in elements likely affected by observational systematics, such as Ag and Cd ($Z = 47, 48$), where strong 3D and NLTE effects are expected, and in third-peak elements sensitive to nuclear-physics inputs such as masses and fission properties. In addition, we perform several tests excluding Ag and Cd, or Ga, Ge, and As, but this leaves the results for $Z > 33$ essentially unchanged.
\end{enumerate}

Beyond inferring conditions, the Bayesian framework presented here can be used to predict the abundances of elements that have not been observed in a given star to guide future observational campaigns. The principal remaining uncertainties are of observational and nuclear-physical rather than astrophysical nature: Across the wide range of neutron-rich conditions explored here, the dominant limitations are the accuracy of the derived stellar abundances and of the nuclear-physics inputs near the peaks. Larger, homogeneously analyzed samples of r-process-enhanced stars, together with improved atomic data and 3D/NLTE corrections, will be essential to disentangle these effects and to accurately determine the astrophysical conditions of the r-process.

\section{Acknowledgements}
We thank Camilla J. Hansen, Marta Molero, and Ian U. Roederer for useful discussions. This work was supported in part by the Deutsche Forschungsgemeinschaft (DFG, German Research Foundation) -- Project-ID 279384907 -- SFB 1245, and the State of Hessen within the Research Cluster ELEMENTS (Project ID 500/10.006). I.S. was supported in part by the European Research Council (ERC) under the European Union's Horizon 2020 research and innovation programme (Grant Agreement No.~101020842).

\bibliography{bibliography}{}
\bibliographystyle{aasjournal}

\end{document}